\documentclass[10pt,twocolumn,letterpaper]{article}

\usepackage{cvpr}              
\definecolor{cvprblue}{rgb}{0.21,0.49,0.74}
\usepackage[pagebackref,breaklinks,colorlinks,allcolors=cvprblue]{hyperref}
\usepackage{url}
\usepackage{booktabs}
\usepackage{amsmath}
\usepackage{graphicx}
\usepackage{xcolor}
\usepackage{algorithm}
\usepackage{subcaption}
\usepackage{placeins}
\newcommand{\Paragraph}[1]{\vspace{1mm} \noindent \textbf{#1} \hspace{0mm}}
\usepackage{booktabs}
\usepackage{adjustbox}
\usepackage{multicol}
\usepackage{bm} 

\newcommand{\vf}{\mathbf{f}}  
\newcommand{\vn}{\mathbf{n}}  
\newcommand{\mL}{\mathbf{L}}  
\newcommand{\vr}{\mathbf{r}}  
\definecolor{DarkRed}{rgb}{0.5,0.1,0.1}
\definecolor{DarkGreen}{rgb}{0.2, 0.5, 0.0}
\definecolor{DarkBlue}{rgb}{0.1, 0.1, 0.9}

\def\paperID{-}
\def\confName{CVPR}
\def\confYear{2026}

\title{OrgFlow: Generative Modeling of Organic Crystal Structures \\ from Molecular Graphs}

\author{Mohammadmahdi Vahediahmar\\
Drexel University, College of Computing and Informatics\\
{\tt\small av834@drexel.edu}
\and
Matthew A. McDonald\\
Drexel University, Department of Chemical and Biological Engineering\\
{\tt\small mam3266@drexel.edu}
\and
Feng Liu\\
Drexel University, College of Computing and Informatics\\
{\tt\small fl397@drexel.edu}
}

\author{
Mohammadmahdi Vahediahmar$^{1}$\quad
Matthew A. McDonald$^{2}$\quad
Feng Liu$^{1}$\\
$^1$ Department of Computer Science, Drexel University\\
$^2$ Department of Chemical and Biological Engineering, Drexel University\\
}

\begin{document}
\maketitle
\begin{abstract}
Crystal structure prediction is a long-standing challenge in materials science, with most data-driven methods developed for inorganic systems. This leaves an important gap for organic crystals, which are central to pharmaceuticals, polymers, and functional materials, but present unique challenges, such as larger unit cells and strict chemical connectivity. We introduce a flow-matching model for predicting organic crystal structures directly from molecular graphs. The architecture integrates molecular connectivity with periodic boundary conditions while preserving the symmetries of crystalline systems. A bond-aware loss guides the model toward realistic local chemistry by enforcing distributions of bond lengths and connectivity. To support reliable and efficient training, we built a curated dataset of organic crystals, along with a preprocessing pipeline that precomputes bonds and edges, substantially reducing computational overhead during both training and inference. Experiments show that our method achieves a Match Rate more than 10 times higher than existing baselines while requiring fewer sampling steps for inference. These results establish generative modeling as a practical and scalable framework for organic crystal structure prediction.
\end{abstract}    
\section{Introduction}
\label{sec:intro}

Crystal structure prediction (CSP) seeks to determine the periodic arrangement of atoms and the unit cell from composition and basic constraints. It is central to materials discovery~\cite{Bowskill2021CSPReview,Zhu2023CSPReview}, drug development and solid-form selection~\cite{Hilfiker2018Polymorphism}, and chemical manufacturing where crystallization governs yield, purity, and scalability~\cite{Myerson2019HandbookCrystallization}. For small-molecule drugs, different polymorphs, which are chemically identical molecules packed into different crystal structures, can shift bioavailability, stability, and shelf life, so anticipating packing early reduces trial-and-error screening and shortens development cycles~\cite{Hilfiker2018Polymorphism}. In this work we focus on the \emph{\textbf{organic-molecule}} setting, where packing-driven property changes are especially consequential.

\begin{figure}[t]
    \centering
    \includegraphics[width=\linewidth]{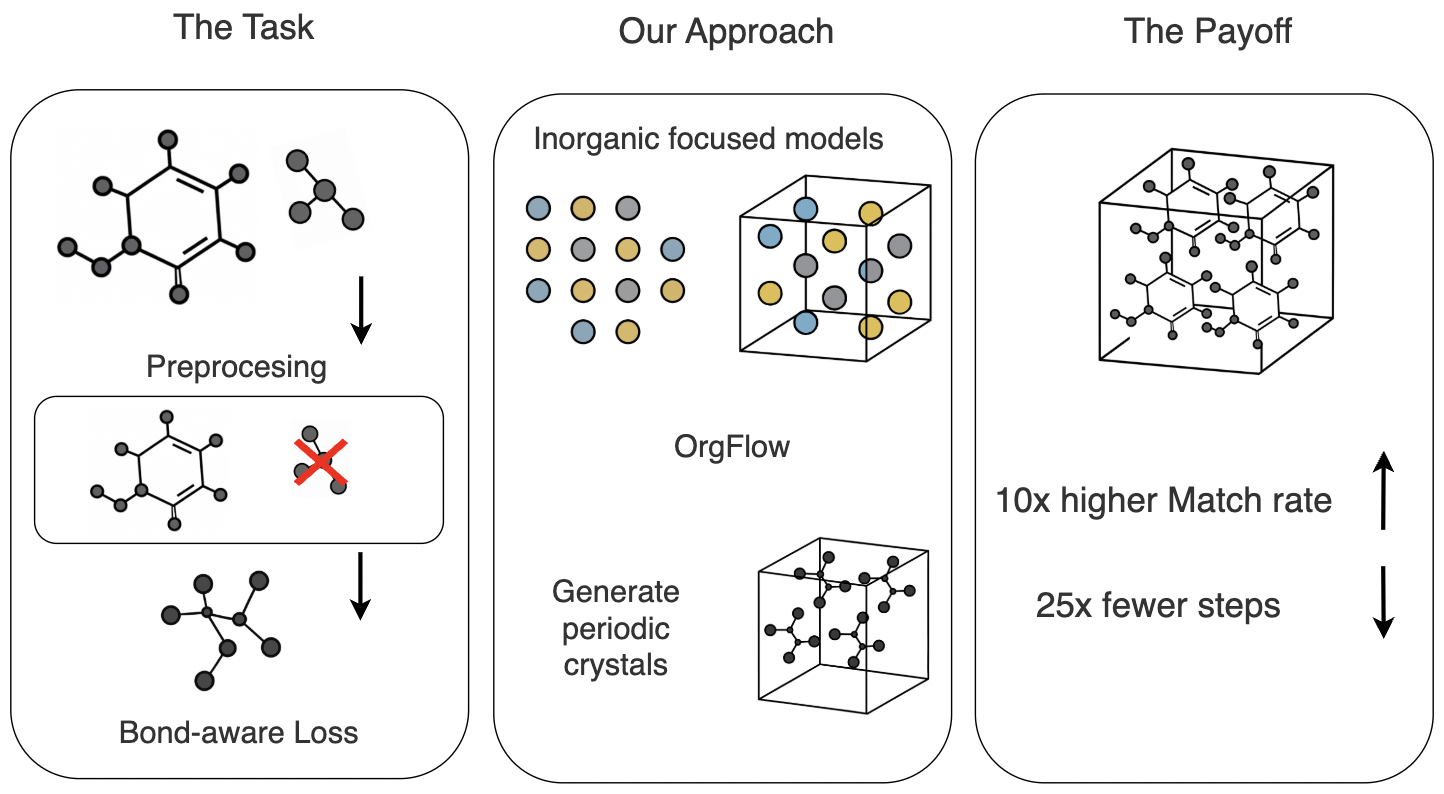}
    \caption{Overview of OrgFlow. OrgFlow learns to generate organic crystal structures from molecular graphs. It preserves covalent bonds through a bond-aware loss, overcoming limitations of inorganic-focused models. The result is accurate, periodic crystals with more than 10× higher match rates and 25× fewer sampling steps.}
    \label{fig:tease}
\end{figure}


Recent years have witnessed an explosion of generative approaches to crystal structure prediction, spanning from diffusion models that progressively denoise lattices and atomic coordinates to flow-matching methods that learn smooth transport between distributions \cite{Xie2021CDVAE,Jiao2023DiffCSP,Lin2024EquiCSP,EkstromKelvinius2025WyckoffDiff,Lipman2023FlowMatching,Liu2023RectifiedFlow,Miller2024FlowMM,Sriram2024FlowLLM,Luo2025CrystalFlow}. Despite these advances, a critical limitation persists: the vast majority of methods target \textbf{\emph{inorganic}} crystals and rely exclusively on inorganic benchmarks for evaluation \cite{Xie2021CDVAE,Jiao2023DiffCSP,Lin2024EquiCSP,EkstromKelvinius2025WyckoffDiff,Miller2024FlowMM,Sriram2024FlowLLM,Luo2025CrystalFlow}. Consequently, the covalent connectivity inherent to organic molecules, a fundamental structural constraint, is rarely leveraged as a conditioning signal. This oversight leaves crucial organic-specific phenomena underexplored, including intramolecular geometry preservation, hydrogen-bond network formation, and $\pi$--$\pi$ stacking interactions. Moreover, the characteristically larger and more conformationally flexible unit cells of organic crystals pose additional challenges that current methods neither explicitly address during training nor adequately assess during evaluation \cite{Bowskill2021CSPReview,Zhu2023CSPReview}. While diffusion-based approaches have shown promise, their computational burden of requiring tens to hundreds of denoising steps significantly limits practical applicability \cite{Xie2021CDVAE,Jiao2023DiffCSP,Lin2024EquiCSP,EkstromKelvinius2025WyckoffDiff}. Flow-based generators offer improved sampling efficiency \cite{Miller2024FlowMM,Sriram2024FlowLLM,Luo2025CrystalFlow}, yet remain predominantly demonstrated on inorganic systems. \emph{These fundamental gaps underscore the urgent need for an organic-focused generator that explicitly conditions on molecular connectivity while maintaining crystallographic symmetry constraints and achieving computationally efficient sampling.}

In light of this, we propose \textbf{OrgFlow}, a conditional flow-matching framework that explicitly incorporates molecular connectivity constraints for organic crystal prediction. Our technical contributions are threefold: (1) We formulate CSP as transport from a Gaussian prior to crystal distributions conditioned on molecular graphs, where edges encode covalent bonds as hard constraints throughout the generation; (2) We design an E(3)-equivariant message-passing architecture operating on periodic graphs with integer image shifts, jointly learning velocity fields for fractional coordinates and lattice parameters while preserving crystallographic symmetries; (3) We introduce a \emph{bond-aware regularization loss} computed from empirical bond-length distributions, enabling chemically valid generation without expensive quantum mechanical calculations. Following rectified flow principles~\cite{Lipman2023FlowMatching,Liu2023RectifiedFlow}, our formulation yields near-linear interpolation paths, achieving convergence in 20 ODE steps, a 25-fold reduction compared to baselines.

Specifically, our pipeline transforms raw crystallographic information files (CIFs) into periodic molecular graphs suitable for generative modeling. We extract molecular connectivity from CIF topology, construct edges with integer lattice shifts for proper boundary handling, and remove poorly-resolved hydrogen atoms for later reconstruction. The equivariant network jointly learns velocity fields for fractional coordinates and lattice parameters, regularized by bond-aware losses computed from empirical distance statistics. 
To support OrgFlow's training, we curate 177k organic structures from the Cambridge Structural Database (CSD)~\cite{Groom2016CSD}, computing bond-length distributions across diverse chemical environments. Our benchmark spans drug-like molecules, small organics, and general pharmaceutical compounds, each presenting distinct packing challenges. Evaluation employs crystallography-aware metrics: pymatgen's StructureMatcher for geometric agreement and spglib for symmetry validation. OrgFlow achieves $21.94\%$ match rate on drug-like molecules versus $0.1\%$ for baselines, validating our molecular conditioning approach.

Although several machine-learning approaches target \emph{molecular} crystals, including DeepCSP~\cite{ye2024organic} and recent ML-guided CSP workflows~\cite{galanakis2024rapid,taniguchi2025crystal}, these systems either rely on template or topological priors or do not natively condition on the molecular graph within a flow-matching generator. To our knowledge, \textbf{OrgFlow} is the first \emph{flow-matching} framework that (i) conditions explicitly on the asymmetric-unit molecular graph, (ii) trains on a large organic-only corpus with molecule-disjoint splits, and (iii) introduces a bond-aware Student-t regularizer tailored to organic covalent geometry.

In summary, the contributions of this work include:

$\diamond$ We propose a novel framework, \textbf{OrgFlow}, that generates full periodic \textbf{\emph{organic}} crystal structures directly from molecular graphs via conditional flow matching.

$\diamond$ We propose a bond-aware regularization loss that enforces chemically valid geometries using empirical statistics, eliminating the need for expensive quantum mechanical validation while improving match rates.

$\diamond$ We establish a large-scale, 177k-structure organic-crystal benchmark derived from the CSD, spanning diverse chemistries and packing motifs, with predefined molecule-disjoint partitions that serve as a valuable community reference for benchmarking organic CSP.

$\diamond$ Extensive experiments show \textbf{OrgFlow} achieves superior performance over recent diffusion- and flow-based baselines.
\section{Related Work}
\label{sec:related}

\Paragraph{Organic Crystal Structure Prediction.}
Organic CSP has been shaped by decades of community benchmarks and industrial validation. The CCDC blind tests standardized performance comparisons for molecular crystals \cite{reilly2016report,hunnisett2024seventh}, and recent large-scale pharmaceutical studies highlight the importance of polymorphism control \cite{zhou2025robust}. Classical search-based frameworks (\eg, USPEX) explore vast packing spaces coupled with energy ranking \cite{glass2006uspex}, but the computational cost scales poorly with molecular flexibility. These limitations motivate learning-based approaches that can infer packing directly from molecular graphs.  
Recent works have started to apply data-driven approaches to organic CSP. For instance, \cite{ye2024organic} introduced DeepCSP, a coupled generative adversarial network and graph convolutional framework that predicts organic crystal lattice parameters within minutes, demonstrating that machine learning can drastically reduce search time; however, it does not predict atomic coordinates or molecular conformations required for further evaluation. Other hybrid pipelines combine statistical models with density functional theory to pre-screen conformers and polymorphs, providing a scalable bridge between empirical heuristics and full \textit{ab initio} simulations.

\Paragraph{Generative Models for Crystals.}
Equivariant neural networks have transformed crystal representation learning. {CGCNN} \cite{xie2018crystal} first modeled periodic graphs for property prediction. Subsequent equivariant models, such as {E(n)-GNN} \cite{satorras2021n}, {SE(3)-Transformer} \cite{fuchs2020se}, and {PaiNN} \cite{schutt2021equivariant}, improved geometric fidelity by enforcing symmetry constraints.  
Generative variants build upon these representations to directly sample periodic structures. CDVAE integrated a diffusion process into a VAE to generate crystal structures with high symmetry validity \cite{Xie2021CDVAE}. Recent diffusion and flow-based models, such as DiffCSP, MatterGen, and FlowMM, jointly generate lattice parameters and atomic coordinates, achieving stable and diverse inorganic structures \cite{Miller2024FlowMM,chen2023flow}.  
More recently, several studies have focused on explicitly encoding crystallographic symmetries into generative architectures. WyCryst \cite{zhu2024wycryst} introduced a Wyckoff-based generative framework that guaranties valid symmetry by operating directly in the space of symmetry operations. WyckoffDiff \cite{EkstromKelvinius2025WyckoffDiff} extends this idea using a diffusion model constrained by Space Group operations, while Space Group Equivariant Diffusion (SGEDiff) \cite{chang2025space} learns Space Group equivariance to maintain consistent symmetry across generated samples. These advances are particularly useful for inorganic materials, which tend to occupy higher-order space groups compared to organic molecules, which occupy groups with fewer symmetry operations owing to their often asymmetrical molecular structures. However, symmetry-aware modeling remains valuable for all crystal structure prediction tasks. 

\Paragraph{Conditioned and Guided Generation}
Conditioned generation aims to produce structures that are consistent with input constraints, such as atomic composition or molecular structure. For crystal systems, conditioning has been applied to enforce space-group symmetries or target-specific compositions \cite{Jiao2023DiffCSP}. 
%
Symmetry-Constrained Diffusion Models (SCDM) extend this principle to 2D materials by incorporating explicit Wyckoff and space-group control during generation, ensuring the crystallographic validity and thermodynamic stability of the generated sheets \cite{xu2025discovery}. However, 
few
 existing methods integrate molecular connectivity or organic-specific priors, which are critical for maintaining realistic covalent geometries in organic crystals. Our work bridges this gap by unifying molecular-graph conditioning with a bond-aware guidance term, ensuring that the generated crystals align with both molecular conformation and periodic lattice constraints. This integration enables realistic organic crystal generation while preserving long-range symmetry.

\begin{figure*}[tbh]
\centering

\begin{subfigure}[t]{0.48\textwidth}
    \centering
    \includegraphics[width=\linewidth]{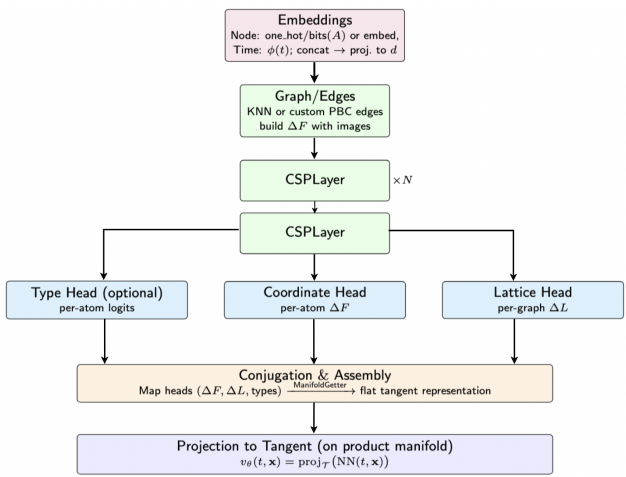}
    \caption{Model architecture. Atomic and time embeddings form graph inputs, periodic edges encode PBCs, and CSP layers predict coordinate and lattice velocities.}
    \label{fig:architecture_overview}
\end{subfigure}
\hfill
\begin{subfigure}[t]{0.48\textwidth}
    \centering
    \includegraphics[width=\linewidth]{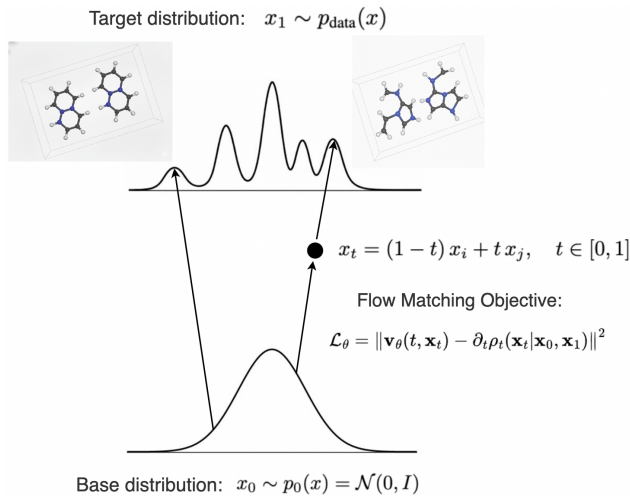}
    \caption{Flow-matching formulation. The model learns a velocity field that transports samples from a base distribution to the crystal data distribution.}
    \label{fig:flow_matching_overview}
\end{subfigure}

\caption{
\textbf{Overview of OrgFlow.} 
\subref{fig:architecture_overview} shows the architecture combining molecular embeddings, periodic edges, and symmetry-preserving layers. 
\subref{fig:flow_matching_overview} illustrates the conditional flow-matching process used to generate full crystal structures.
}
\label{fig:combined_overview}
\end{figure*}

\section{Methodology}
\label{sec:methodology}

\subsection{Problem Formulation}
We pose crystal structure prediction (CSP) as a conditional generative modeling problem:
\[
X \sim p_\theta(X \mid G),
\]
where $G$ is the molecular graph of the asymmetric unit and $X = (\mathbf{R}, \mathbf{L})$ denotes the crystal structure defined by atomic fractional coordinates $\mathbf{R} \in [0,1)^{n\times3}$ and the lattice matrix $\mathbf{L} \in \mathbb{R}^{3\times3}$. The objective is to learn a generative model that transforms samples from an easily sampled Gaussian distribution into realistic crystal configurations conditioned on $G$. See \Cref{fig:combined_overview} for an illustration of the model architecture and its overall workflow. Empirically, high-density regions of the learned distribution tend to correspond to low-energy minima in the crystal energy landscape.  
Additionally, we explicitly minimize covalent bond strain to encourage chemically valid molecular conformations. This dual objective aligns both global crystal packing and local molecular realism.

Formally, the generative process is modeled through a time-dependent vector field $\mathbf{f}_\theta(X_t, t, G)$ that transports $X_0 \sim p_0(X)$ toward $X_1 \sim p_{\text{data}}(X \mid G)$ via the continuity equation:
\[
\frac{\partial p_t(X)}{\partial t} + \nabla_X \cdot \left(p_t(X)\mathbf{f}_\theta(X_t, t, G)\right) = 0.
\]
This formulation allows efficient, deterministic sampling compared to diffusion models while ensuring smooth evolution between base and target distributions.

\subsection{Preprocessing}
We convert each CIF entry into a symmetry-consistent crystal graph suitable for conditional CSP. The pipeline preserves periodicity, encodes custom molecular connectivity, and attaches the bond statistics required by our bond-aware loss.  
Currently, we focus on crystals with a single asymmetric unit ($Z' = 1$) and integer copies of that unit in the cell ($Z = 1,2,3,\dots$).

\Paragraph{Input normalization and symmetry consideration.}
We parse the unit cell, species, and fractional coordinates from CIF files representing structures determined by X-ray crystallography from the CSD. We apply standard reductions (e.g., Niggli) and space-group refinement to obtain a canonical, symmetry-consistent cell. When symmetry operations are available, we verify them against the refined structure and record both the space-group number and per-site Wyckoff information for later use in validation and augmentation.

\Paragraph{Molecular validity and connectivity.}
If a SMILES string is provided for the asymmetric unit, we first confirm chemical validity and restrict atom types to a curated organic set (H, B, C, N, O, F, Si, P, S, Cl, Br, I). We then ensure that molecular connectivity forms a single component using a standard Union–Find algorithm:
\[
\begin{aligned}
\mathrm{find}(i) &=
\begin{cases}
i & P[i]=i,\\
\mathrm{find}(P[i]) & \text{otherwise.}
\end{cases} \\
\mathrm{union}(i,j) &:\; P[\mathrm{find}(i)] \leftarrow \mathrm{find}(j)
\end{aligned}
\]

Disconnected fragments are discarded. Hydrogen atoms, which constitute roughly half of all atoms but are poorly resolved in X-ray crystallography, are removed to reduce graph size. The positions of non-labile hydrogen atoms can be reconstructed from the heavy-atom geometry.

\Paragraph{Custom connectivity under periodic boundary conditions.}
A molecule need not reside entirely within one symmetric unit (the unit cell), therefore, we lift the molecular bonds into the periodic crystal lattice. Let $\vf_i,\vf_j \in [0,1)^3$ denote fractional coordinates and $\mL$ the lattice matrix. We connect $(i,j)$ with a periodic edge and integer image shift $\vn \in \mathbb{Z}^3$ under the nearest-image convention:
\[
\vn = \mathrm{round}(\vf_j - \vf_i), \qquad
\vr_{ij} = \mL\,(\vf_j - \vf_i + \vn),
\]
ensuring distances are computed consistently across cell boundaries.  
To capture short and medium-range interactions, we consider a $3\times3\times3$ neighborhood of image offsets. Duplicate periodic edges are merged, true self-loops are removed, and all edge indices are validated.

\Paragraph{Periodic crystal graph construction.}
If explicit connectivity is unavailable, we construct periodic neighbor graphs using a crystal-aware local environment method. Each edge stores atom indices and an integer image shift, making periodic boundary condition (PBC) distance calculations unambiguous:
\[
\|\vr_{ij}\|_2 = \big\|\mL\,(\vf_j - \vf_i + \vn)\big\|_2,
\]
which directly supports our bond-aware loss and flow-matching objective.

\Paragraph{Bond orders and statistics.}
For each bonded atom pair, we assign a discrete bond order from \{SINGLE, DOUBLE, TRIPLE, AROMATIC\}. Each edge stores the empirical bond-length mean $\mu_{uv}$ and variance $\sigma_{uv}^2$, computed from dataset-wide atom-type distributions. Rare bond types are ignored because a reliable mean and variance are not available. These attributes remain fixed throughout training and inference, ensuring consistent use of bond priors.

\Paragraph{Quality and sanity checks.}
We verify that the periodic graph is non-empty, all indices are valid, and atomic numbers fall within expected ranges. Space-group and Wyckoff metadata are preserved. Any entry failing validation is excluded, yielding a PBC-consistent, symmetry-aware crystal graph with reliable bond annotations, suitable for conditional CSP training.

\subsection{Neural Network Architecture}

Our model is based on equivariant graph neural networks (EGNNs) \cite{satorras2021n}, which ensure translation invariance and rotation equivariance. We adopt the periodic adaptation proposed by \citet{Jiao2023DiffCSP}, which modifies EGNN message passing to handle periodic image shifts.

Compared to FlowMM \cite{Miller2024FlowMM}, which employs large hidden dimensions, we reduce the hidden size from 512 to 128 while increasing the number of layers from 6 to 12. Additionally, a time embedding dimension of 256 and the SiLU activation function were used. These hyperparameters were determined by sweeping parameter values to balance precision and computational cost. This configuration yields a sixfold smaller model without measurable performance loss. The lightweight design accelerates both training and inference while preserving stability and accuracy.

The network outputs velocity fields for both atomic fractional coordinates and lattice parameters, which define the instantaneous flow $\mathbf{f}_\theta(X_t, t, G)$ used during the integration of the flow-matching ODE.

\subsection{Hyperparameters}
Training hyperparameters are tuned separately for different dataset splits.  
$\lambda_f$ balances the fractional coordinate loss, stabilizing atom position optimization within the unit cell. $\lambda_l$ weights the lattice component, and $\lambda_b$ scales the bond-length loss. Selective annealing strategies further improve optimization stability.  
The complete dataset-specific hyperparameters are listed in \cref{tab:dataset_hyper}.


\begin{table*}[tb]
\centering
\caption{Training hyperparameters for different organic crystal datasets.}
\label{tab:dataset_hyper}
\begin{tabular}{lcccc}
\toprule
 & Drug-like & Small-molecules & Smallest-molecules & Non-drug-like \\
\midrule
Learning Rate & 0.0003 & 0.0001 & 0.0001 & 0.0003 \\
Weight Decay & 0.0001 & 0.0001 & 0.0001 & 0.0 \\
$\lambda_f$ (Frac Coords) & 600 & 400 & 400 & 600 \\
$\lambda_l$ (Lattice) & 1 & 1 & 1 & 1 \\
$\lambda_b$ (Bond Length) & 0.002 & 0.0015 & 0.001 & 0.001 \\
Anneal f & True & True & True & True \\
Anneal l & False & False & False & False \\
\bottomrule
\end{tabular}
\end{table*}

\subsection{Bond-aware Loss}
We regularize predictions using a bond-length prior derived from empirical distributions. 
For each edge $(u,v)$ with a predicted interatomic distance $d_{uv}$ and bond statistics $(\mu_{uv}, \sigma_{uv})$, we define:
\[
\mathcal{L}_{\text{bond}} = \frac{1}{|E|}\sum_{(u,v)} \Bigg[ 
2\log\Big(1+\frac{1}{2}\Big(\frac{d_{uv}-\mu_{uv}}{\sigma_{uv}}\Big)^2\Big)
+ \log\sigma_{uv}
\Bigg]
\]
This Student-t penalty uses a fixed degree of freedom of 3. It remains fully differentiable and robust to outliers, and integrates seamlessly with the flow-matching objective. It encourages physically plausible bond geometries and penalizes stretched or compressed covalent bonds that violate chemical constraints.

\Paragraph{Overall Objective.}
The total loss function is:
\[
\mathcal{L}_{\text{flow}} = \lambda_f.\mathcal{L}_{\text{frac}} + \lambda_l.\mathcal{L}_{\text{lattice}} + \lambda_b.\mathcal{L}_{\text{bond}},
\]

The $\mathcal{L}_{\text{frac}}$ term constrains atomic fractional coordinates to remain consistent with the real-space distributions seen in experimentally resolved structures, while $\mathcal{L}_{\text{lattice}}$ enforces alignment of the predicted unit-cell geometry with the ground truth. Together, these components balance local chemical validity and global crystallographic accuracy, ensuring that the generated structures are both physically plausible and symmetrically consistent.

\section{Evaluation}
\label{sec:evaluation}

We evaluate our model based on its ability to reconstruct experimentally determined organic crystal structures. Our main metric is the \textbf{Match Rate}: the fraction of predictions that match the ground truth under a defined tolerance. Matching is performed with \texttt{StructureMatcher} from \texttt{pymatgen}, which aligns lattices under periodic boundary conditions and compares atomic positions using RMSD and tolerance thresholds. This provides a clear measure of geometric fidelity and is standard in CSP benchmarks. OrgFlow and FlowMM use identical lattice and site tolerances following the settings in FlowMM~\cite{Miller2024FlowMM}, and we report results consistently across all splits.

\subsection{Baseline parity and adaptations.}
To ensure a fair comparison, we adapted \textbf{FlowMM}~\cite{Miller2024FlowMM} to load our organic dataset format and periodic molecular graphs. We matched \emph{hardware}, \emph{batch size}, and \emph{numeric precision} across all runs. Specifically, both OrgFlow and FlowMM were trained and evaluated on identical GPUs with the same global batch size and \texttt{float32} precision. Beyond data loading, we tuned the FlowMM learning rate and weight decay on a held out validation subset of the organic dataset while keeping its architecture and loss terms identical to those of the original work. OrgFlow and FlowMM were trained for the same number of optimization steps, used the same set of random seeds, and we applied the ODE integrator and step schedule recommended in the FlowMM paper when evaluating the baseline. Any remaining FlowMM preprocessing changes were limited to data adapters and did not modify model capacity or objectives, so gains reflect the conditional formulation and bond-aware loss rather than extra compute or more favorable settings for OrgFlow.

\Paragraph{Metric interpretation.}
A higher match rate indicates that predicted structures are geometrically and symmetrically consistent with their experimental counterparts. Because organic crystals often exhibit multiple polymorphic forms, not all mismatched predictions represent true model failures, some may correspond to unobserved but physically valid polymorphs. 

\subsection{Dataset}
\label{sec:dataset}

Raw data were obtained from the Cambridge Structural Database (CSD) hosted by the Cambridge Crystallographic Data Center (CCDC)\footnote{\url{https://www.ccdc.cam.ac.uk/}}. Each entry corresponds to a crystal structure determined by single-crystal X-ray diffraction, paired with metadata and curated connectivity annotations. We define four evaluation splits. Drug-like contains the top 20\% of molecules ranked by the Quantitative Estimate of Drug-likeness (QED $\geq 0.73$), while Not Drug-like contains the bottom 20\% (QED $\leq 0.33$). Small-molecule includes molecules with fewer than 24 heavy atoms, and Smallest-molecule includes molecules with fewer than 18 heavy atoms, so the splits probe both drug-likeness and molecular size. 
.

\Paragraph{Preprocessing scripts and data availability.}
We release the full preprocessing pipeline that rebuilds our training triples from raw CSD exports: CIF parsing and refinement, SMILES validation, symmetry handling, periodic edge construction with integer image shifts, bond-statistics extraction, and split assignment. The repository contains end-to-end scripts to regenerate all intermediate artifacts used by OrgFlow and FlowMM in this paper. Due to CSD licensing, we do not redistribute raw CIFs. Our repository provides scripts that reproduce all derived artifacts from a licensed CSD installation. The CCDC allows limited access to CSD data online without a license through WebCSD.
\begin{table}[h]
\centering
\caption{Match Rate across dataset splits.}
\label{tab:split}
\resizebox{1\linewidth}{!}{
\begin{tabular}{lccc}
\toprule
Dataset & FlowMM~\cite{Miller2024FlowMM} & OrgFlow & Increase(\%) \\
\midrule
Whole dataset & $0.4$ & $\mathbf{10.7}$ & $+10.3$ \\
Drug-like & $0.1$ & $\mathbf{21.94}$ & $+21.84$ \\
Not Drug-like & $0.15$ & $\mathbf{15.43}$ & $+15.28$ \\
Small-molecule & $0.43$ & $\mathbf{17}$ & $+16.57$ \\
Smallest-molecule & $0.5$ & $\mathbf{24.55}$ & $+24.05$ \\
\bottomrule
\end{tabular}
}
\end{table}

Table~\ref{tab:split} shows that OrgFlow substantially outperforms FlowMM across all splits, and the magnitude of these gains helps clarify why organic CSP is a fundamentally different setting from the inorganic-focused benchmarks for which FlowMM was designed. FlowMM relies on learning a one-to-many generative map from atomic lattices alone, without conditioning on molecular connectivity. This works reasonably well for inorganic crystals but collapses on organics, where the asymmetric unit contains flexible molecules with dozens of degrees of freedom and chemically meaningful bond geometry that cannot be inferred from lattice information alone. OrgFlow directly conditions on the molecular graph, which constrains the generative path and prevents unrealistic distortions during integration. This explains both the much higher match rates and the fact that OrgFlow’s performance is more stable across splits: trends in QED or molecular size change the difficulty of the task, but they do not reduce the usefulness of graph conditioning. FlowMM, by contrast, has no access to such structure, so its performance varies inconsistently with drug-likeness or heavy-atom count. Overall, these results highlight the intrinsic difficulty of reconstructing full periodic structures from molecular graphs and show that conditioning and bond-aware priors are essential for achieving non-trivial accuracy in organic CSP.

\Paragraph{Dataset bias and interpretation.}
The low match rate observed in baseline models partly stems from incomplete structural coverage rather than inherent model limitations. Organic crystals exhibit polymorphism, the ability of a single molecule to form multiple stable crystal arrangements. 
For most molecules, only a single polymorph is reported. Pharmaceutical compounds, which are heavily studied, often have multiple known structures, which explains the higher match rates for the drug-like subset.  
Conversely, molecules that have crystallized only once often lack alternative structures, making the generated predictions of plausible unobserved polymorphs unmatched.




\subsection{Qualitative Evaluation}
\begin{figure*}[t]
    \centering
    \begin{subfigure}[b]{0.32\textwidth}
        \centering
        \includegraphics[width=\linewidth]{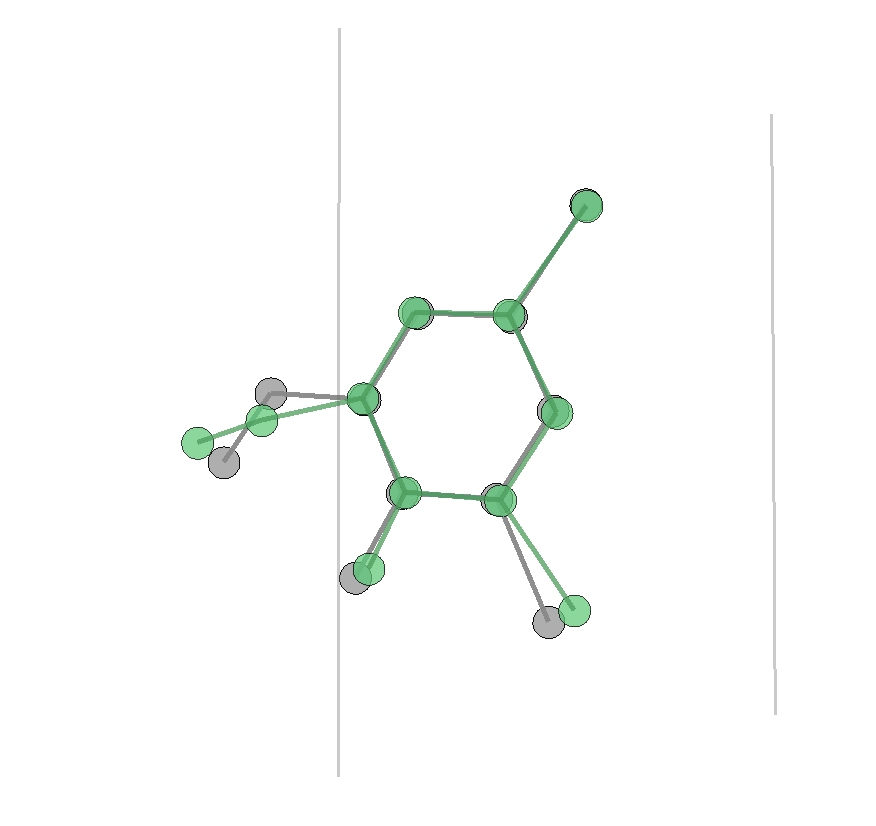}
        \caption{}
        \label{fig:rmsd33}
    \end{subfigure}
    \begin{subfigure}[b]{0.32\textwidth}
        \centering
        \includegraphics[width=\linewidth]{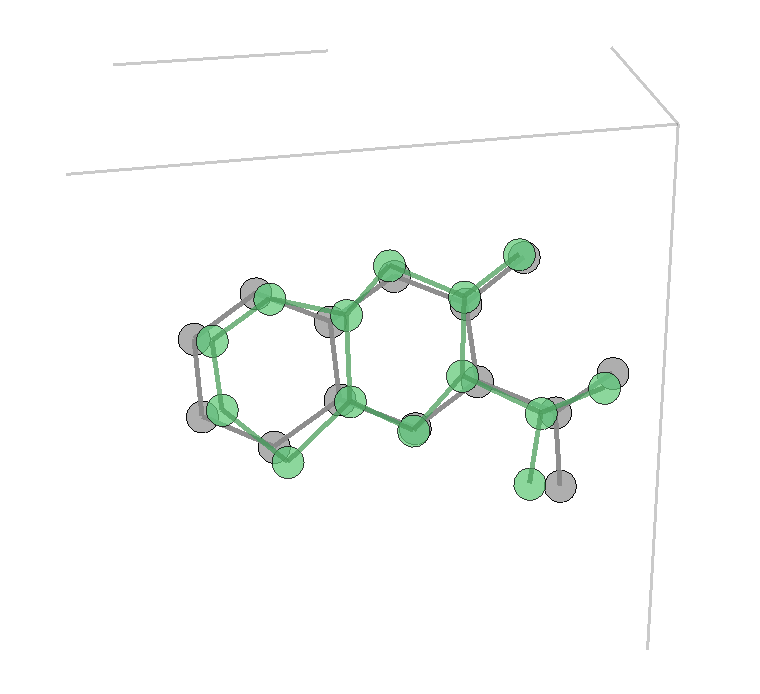}
        \caption{}
        \label{fig:rmsd3579}
    \end{subfigure}
    \begin{subfigure}[b]{0.32\textwidth}
        \centering
        \includegraphics[width=\linewidth]{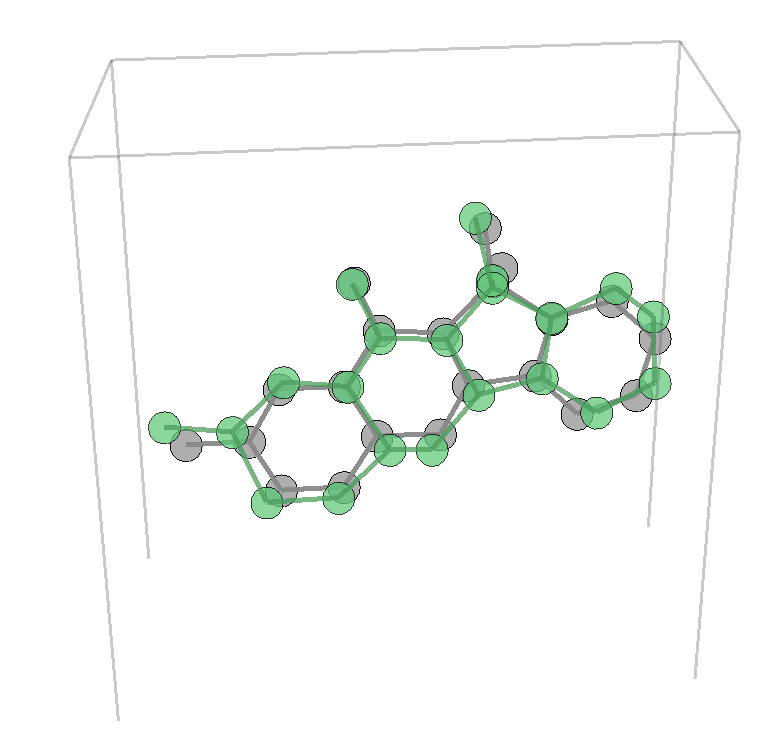}
        \caption{}
        \label{fig:rmsd40}
    \end{subfigure}
    \caption{Examples of predicted crystal structures (green) compared to ground truth (gray). The figures show two-dimensional projections of three-dimensional structures. Only heavy atoms are shown for clarity.}
    \vspace{-2mm}
    \label{fig:pred_vs_gt}
\end{figure*}

Figure~\ref{fig:pred_vs_gt} visualizes predicted versus ground truth crystal structures. Green atoms represent model predictions, and gray atoms denote the reference structures. These comparisons show close agreement in both atomic positions and lattice geometry, validating the model’s ability to recover experimental configurations.

Figure~\ref{fig:pred_vs_gt} highlights the typical behavior of OrgFlow on organic crystals. The model closely reproduces cyclic cores and aromatic ring systems, including ring geometry and relative orientations, and it recovers the overall lattice shape and molecular packing of the experimental structures. Most discrepancies arise in flexible substituents or terminal functional groups, where torsion angles differ and side chains bend or rotate relative to the ground truth. We also observe small shifts in aromatic stacking distances in some cases, suggesting that long-range packing interactions remain more difficult to match exactly than local intramolecular geometry.
\section{Ablation Study}
\label{sec:ablation}

We started with the question of whether a method designed for inorganic CSP could simply be retrained on organic crystals. Directly adapting FlowMM provides a powerful baseline on inorganics; however, on organic data, it yields low match rates and requires hundreds of sampling steps. This motivated us to design OrgFlow so that it samples more efficiently and encodes organic-specific geometric structure. In this section, we use the \textbf{Small Molecule} split to unpack how each design choice contributes. Each configuration is trained with three seeds, and we report the mean Match Rate and standard deviation.

\Paragraph{Inference efficiency.}

\begin{figure}[htb!]
    \centering
    \begin{subfigure}[b]{0.45\linewidth}
        \centering
        \includegraphics[width=\linewidth]{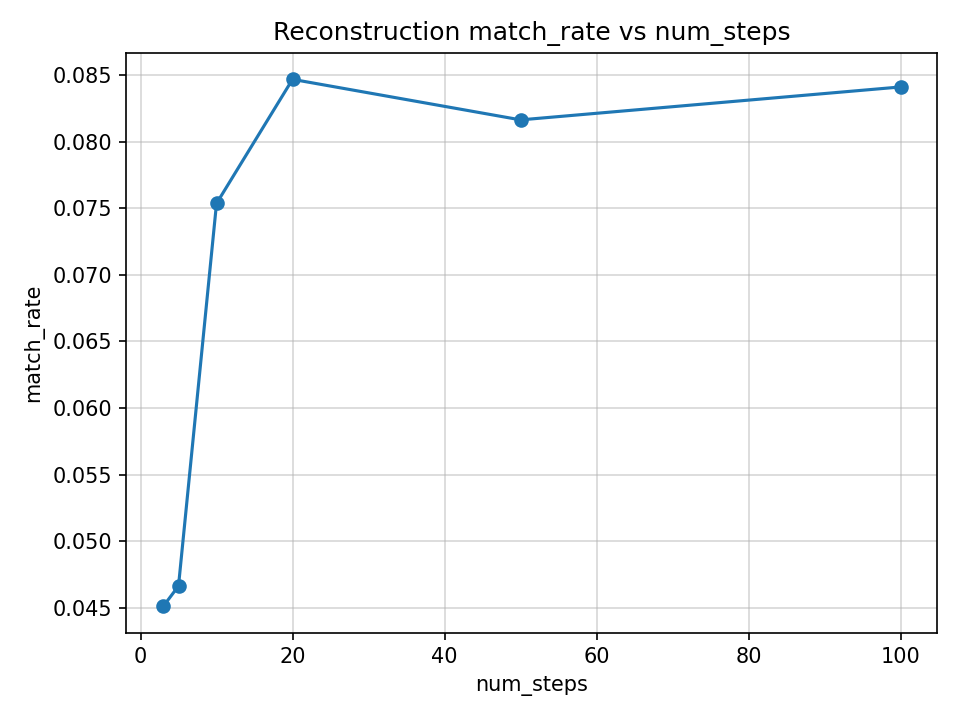}
        \caption{OrgFlow.}
        \label{fig:eff_org}
    \end{subfigure}
    \hfill
    \begin{subfigure}[b]{0.45\linewidth}
        \centering
        \includegraphics[width=\linewidth]{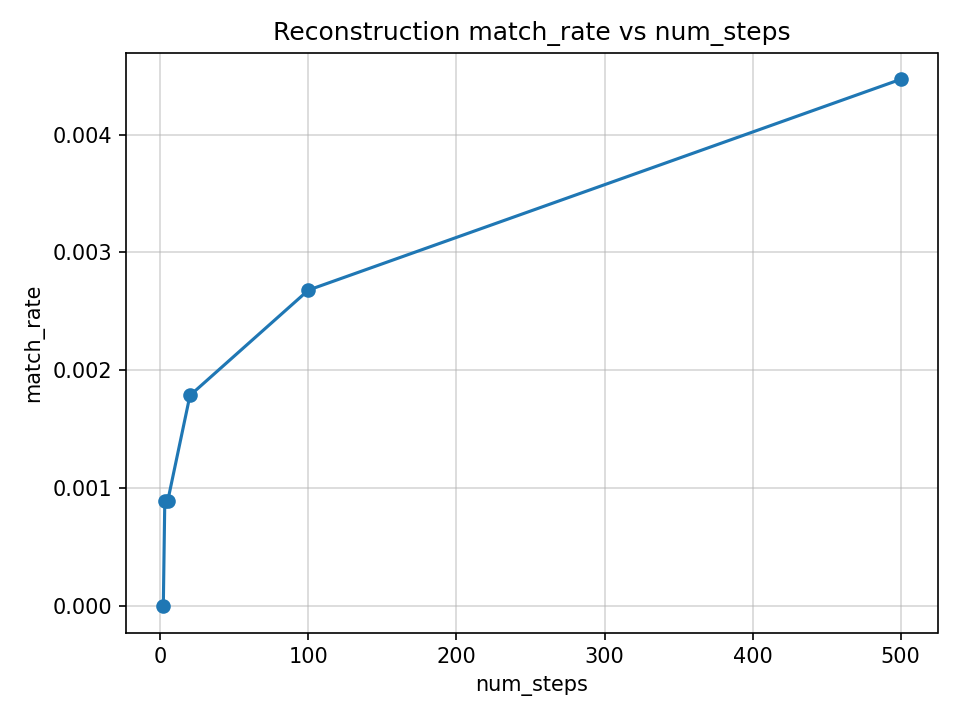}
        \caption{FlowMM.}
        \label{fig:eff_flow}
    \end{subfigure}
    \caption{Inference efficiency on Small Molecules. OrgFlow saturates at about $20$ steps, while FlowMM needs around $500$ to approach its best Match Rate.}
    \label{fig:efficiency_comparison}
\end{figure}

We first compare how quickly the two models utilize their learned transport fields at test time. \Cref{fig:eff_org} and \Cref{fig:eff_flow} show the Match Rate as a function of the number of ODE steps. On the Small Molecule split, OrgFlow reaches near-peak performance with about $20$ steps, while FlowMM needs roughly $500$ steps to achieve a similar Match Rate. This experiment shows that the conditional flow field learned by OrgFlow can move samples toward valid organic packings with fewer integration steps. In other words, once we condition on the molecular graph and let the model directly regress a deterministic vector field, we can trade long trajectories for much shorter transports without sacrificing accuracy. We also measure inference cost by recording wall-clock time and computing seconds per sample as a function of the number of ODE steps. OrgFlow is consistently faster than FlowMM across all shared settings, demonstrating a clear efficiency advantage in addition to higher accuracy. 

\Paragraph{Effect of bond-aware loss.}
A key structural difference between inorganic and organic crystals is that organic molecules have well defined covalent geometries. Naively retraining an inorganic model on organic data does not enforce these local constraints, and we often observe distorted bonds even when the global periodicity is correct. To address this, we introduced a bond-aware loss $\mathcal{L}_{\text{bond}}$ that penalizes deviations from empirical bond-length statistics. The lookup table used in this loss is computed once from the training set (refer to Appendix)
 and reused in all runs.

\Cref{tab:sweep_bl} reports the effect of varying the bond-loss coefficient $\lambda_b$. The Match Rate increases from $6.7\%$ with no bond term to $10.1\%$ at $\lambda_b = 5\times10^{-3}$, with a roughly monotonic gain across the range. We did not push to larger values because training became unstable beyond $\lambda_b = 5\times10^{-3}$. Together, these results show that the bond prior consistently improves performance by correcting local distortions, but it must be kept at a moderate weight so that its gradients do not dominate the optimization.

\begin{table}[tbh!]
\centering
\caption{Effect of bond loss coefficient $\lambda_b$ on Match Rate.}
\label{tab:sweep_bl}
\begin{tabular}{ccc}
\toprule
$\lambda_b$ & Match Rate (\%) & Std \\
\midrule
0.0          & 6.7  & 0.15 \\
5$\times$10$^{-4}$ & 8.2  & 0.18 \\
1$\times$10$^{-3}$ & 8.4  & 0.20 \\
2$\times$10$^{-3}$ & 8.5  & 0.22 \\
5$\times$10$^{-3}$ & 10.1 & 0.25 \\
\bottomrule
\end{tabular}
\vspace{-2mm}
\end{table}

\Paragraph{Effect of fractional coordinate loss.}
Once bond geometries are regularized, the next question is how strongly to penalize errors in the fractional coordinates themselves. The fractional loss $\mathcal{L}_{\text{frac}}$ anchors atoms to realistic positions within the unit cell and helps resolve local packing ambiguities that are not fully determined by bond lengths alone. Keeping $\lambda_b = 1\times10^{-3}$ and $\lambda_l = 1$ fixed, we sweep $\lambda_f$ and report the results in \Cref{tab:sweep_frac}.

The Match Rate rises from $8.7\%$ at $\lambda_f = 200$ to $11.7\%$ at $\lambda_f = 800$, with steady improvements across the range. This suggests that emphasizing fractional-coordinate accuracy complements the bond term: once local bond geometry is reasonable, tightening the distribution of fractional positions helps the model place molecules into more realistic packings. The gains begin to taper at higher $\lambda_f$, which is consistent with the idea that, beyond a point, additional weight on $\mathcal{L}_{\text{frac}}$ mainly refines structures rather than qualitatively changing them.

\begin{table}[h!]
\centering
\caption{Effect of fractional coordinate weight $\lambda_f$ ($\lambda_b = 1\times10^{-3}$, $\lambda_l = 1$).}
\label{tab:sweep_frac}
\begin{tabular}{ccc}
\toprule
$\lambda_f$ & Match Rate (\%) & Std \\
\midrule
200 & 8.7 & 0.21 \\
400 & 9.8 & 0.19 \\
600 & 10.1 & 0.22 \\
800 & 11.7 & 0.13 \\
\bottomrule
\end{tabular}
\end{table}

\Paragraph{Effect of lattice loss.}
Finally, we examine how strongly to constrain the lattice itself. The lattice loss $\mathcal{L}_{\text{lattice}}$ aligns predicted unit cells with ground truth and is necessary for crystallographic validity; however, it interacts with both bond and fractional terms. With $\lambda_b = 1\times10^{-3}$ and $\lambda_f = 600$, we vary $\lambda_l$ as shown in \Cref{tab:sweep_lat}.

Raising $\lambda_l$ from $0.5$ to $4.0$ improves the Match Rate from $9.6\%$ to $13.6\%$, but increasing it to $6.0$ drops performance to $6.7\%$. This suggests that moderate lattice supervision helps the model learn the cell, while stronger penalties force global changes faster than local neighborhoods can adjust. In that setting, $\mathcal{L}_{\text{lattice}}$ starts to conflict with $\mathcal{L}_{\text{bond}}$ and $\mathcal{L}_{\text{frac}}$ rather than supporting them.

\begin{table}[h!]
\centering
\caption{Effect of lattice weight $\lambda_l$ ($\lambda_b = 1\times10^{-3}$, $\lambda_f = 600$).}
\label{tab:sweep_lat}
\begin{tabular}{ccc}
\toprule
$\lambda_l$ & Match Rate (\%) & Std \\
\midrule
0.5 & 9.6 & 0.16 \\
1.0 & 10.1 & 0.15\\
2.0 & 11.0 & 0.18\\
4.0 & 13.6 & 0.21\\
6.0 & 6.7 & 0.12\\
\bottomrule
\end{tabular}
\vspace{-2mm}
\end{table}

\noindent\emph{Why do the coefficients operate on different scales.}
These patterns also explain why the three geometry terms use different numerical scales. Fractional coordinates lie in $[0,1)$ with low variance, so $\mathcal{L}_{\text{frac}}$ needs a larger weight to matter in the total gradient. Lattice parameters are fewer and larger in magnitude, so a modest $\lambda_l$ is enough to shape the cell. The bond term is a heavy-tailed Student-t log-likelihood and already produces strong gradients near outliers, so $\lambda_b$ must remain small to keep training stable. The chosen coefficients balance the gradient magnitudes so that no single term dominates learning.

\Paragraph{Ablation study summary.}
Starting from a naive inorganic-style model, we gain efficiency by learning a conditional flow that requires only tens of steps during inference. We then improve chemical validity by adding a bond-aware prior, refining local packing with a stronger fractional-coordinate loss, and aligning the periodic cell with a moderately weighted lattice term. Best performance arises when these three geometric losses are balanced: the bond prior enforces realistic covalent geometry, the fractional term sharpens atomic positions, and the lattice term corrects global structure without overwhelming local adjustments.
\section{Conclusion}
\label{sec:conclusion}

We introduced \textbf{OrgFlow}, a flow-matching framework for organic crystal structure prediction that conditions on molecular graphs while preserving periodic boundary conditions. By combining molecular connectivity with bond-aware regularization, OrgFlow bridges local chemical validity and long-range crystallographic order, substantially outperforming existing inorganic-focused methods on organic crystals while requiring fewer integration steps than baselines. The bond-aware loss ensures chemically valid geometries without expensive quantum calculations. Many unmatched predictions appear chemically sound but are absent from databases, suggesting undiscovered polymorphs rather than failures. Current limitations in flexible groups highlight opportunities for additional geometric constraints. Future work will extend to multi-component crystals and energy-guided refinement. This establishes molecular graph conditioning as essential for organic crystal structure prediction, providing a foundation for accelerating pharmaceutical development and materials discovery.

{
    \small
    \bibliographystyle{ieeenat_fullname}
    \bibliography{main}
}


\end{document}